\documentclass[10pt,letterpaper]{article}
\usepackage[top=0.85in,left=2.75in,footskip=0.75in]{geometry}

\usepackage{amsmath,amssymb}

\usepackage{changepage}

\usepackage[utf8x]{inputenc}

\usepackage{textcomp,marvosym}

\usepackage{cite}

\usepackage{nameref,hyperref}

\usepackage[right]{lineno}

\usepackage{microtype}
\DisableLigatures[f]{encoding = *, family = * }

\usepackage[table]{xcolor}
\hypersetup{
    colorlinks = true,                    
           }

\usepackage{array}

\newcolumntype{+}{!{\vrule width 2pt}}

\newlength\savedwidth

\usepackage[aboveskip=1pt,labelfont=bf,labelsep=period,justification=raggedright,singlelinecheck=off]{caption}

\makeatletter
\renewcommand{\@biblabel}[1]{\quad#1.}
\makeatother

\usepackage{lastpage,fancyhdr,graphicx}
\usepackage{epstopdf}
\fancyheadoffset[L]{2.25in}
\fancyfootoffset[L]{2.25in}
\begin{document}
\vspace*{0.2in}

\begin{flushleft}
{\Large
\textbf\newline{When the goal is to generate a series of activities: \\
A self-organized simulated robot arm} 
}
\newline
\\
Tim Koglin\textsuperscript{1},
Bulcs\'u S\'andor\textsuperscript{2*},
Claudius Gros\textsuperscript{1}
\\
\bigskip
\textbf{1} Institute for Theoretical Physics, Goethe University Frankfurt, Frankfurt am Main, Germany
\\
\textbf{2} Department of Physics, Babeș-Bolyai University,
Cluj-Napoca, Romania
\\
\bigskip

%
%





*bulcsu.sandor[hereWeAre]phys.ubbcluj.ro (BS)

\end{flushleft}
\section*{Abstract}
Behavior is characterized by sequences of goal oriented 
conducts, such as food uptake, socializing and resting.
Classically, one would define for each task a corresponding
satisfaction level, with the agent engaging, at a given time,
in the activity having the lowest satisfaction level. 
Alternatively, one may consider that the agent follows the
overarching objective to generate sequences of distinct 
activities. To achieve a balanced distribution of activities 
would then be the primary goal, and not to master a specific
task. In this setting the agent would show two types of 
behaviors, task-oriented and task-searching phases, with 
the latter interseeding the former.
We study the emergence of autonomous task switching 
for the case of a simulated robot arm. Grasping one 
of several moving objects corresponds in this 
setting to a specific activity. Overall, the arm should
follow a given object temporarily and then move away,
in order to search for a new target and reengage. We show that 
this behavior can be generated robustly when modeling
the arm as an adaptive dynamical system. The dissipation
function is in this approach time dependent. The arm
is in a dissipative state when searching for a nearby
object, dissipating energy on approach. Once close,
the dissipation function starts to increase, with the
eventual sign change implying that the arm will take 
up energy and wander off. The resulting explorative
state ends when the dissipation function becomes again 
negative and the arm selects a new target. We believe 
that our approach may be generalized to generate 
self-organized sequences of activities in general.




\section*{Introduction}
Besides their industrial and practical applications,
real and simulated robots are used increasingly to study 
the principles underlying embodied cognition
\cite{martius2013information} and locomotion \cite{ijspeert2014biorobotics},
together with the self organization of critical sensorimotor states 
\cite{aguilera2015self} and motor primitives \cite{tani2003self}.
Simulated robots may be considered in addition as proxies for cognitive
and information processing agents \cite{beer2015information}.

It is well known that gaits and other regular muscle contractions, 
like breathing \cite{arshavsky2016central}, are induced in many 
cases by central pattern generators \cite{marder2001central,ijspeert2008central}, 
even though it is currently controversial whether this is the
case for biped locomotion \cite{minassian2017human}, viz for 
human walking. Abstracting from animal models, one may ask conversely to which
extent compliant locomotion may be generated via self-organizing
principles \cite{sandor2015sensorimotor}, that is in the absence 
of top-down control in the form of a central pattern generator.
One talks in this context of `embodiment' \cite{pfeifer2007self},
when part of the computation generating locomotion is carried
out by the elasto-mechanical properties of the constituting body
\cite{aguilar2016review}. For quadruped robots with legs that
are independently controlled by single non-linear phase
oscillators \cite{owaki2013simple}, it has been shown that
the limb-specific sensorimotor feedback derived form pressure 
sensors leads to self-organized interlimb communications, with 
emerging gaits that correspond to walking, trotting and
galloping \cite{owaki2017quadruped}.  

Self-organizing principles may be implemented within the
sensorimotor loop \cite{sandor2015sensorimotor}, which 
is comprised of environment, body, actuator and sensory 
readings, with the latter being restricted in the pure case 
to propiosensation, viz to the internal state of the robot.
The attractors self-stabilizing in the sensorimotor loop
may then give rise to complex patterns of regular and of 
chaotic motion primitives \cite{martin2016closed}, which 
can be selected in a second step using `kick control' 
\cite{sandor2018kick}. From a general perspective, kick 
control is an instance of a higher-level control mechanism 
exploiting the reduction in control complexity provided by
morphologically computing robots
\cite{muller2017morphological,ghazi2017morphological}.
These approaches are hence different from other works 
where closed-loop policies are applied on the top of 
open-loop gait cycles
\cite{sprowitz2013towards,travers2016dynamical}. 
Alternatively, sequential switching between self-organizing behaviors 
in the combined phase space of the controller, body and 
environment can also be generated via 
self-exploration of the attractor landscape using an 
adaptive repelling potential \cite{pinneri2018systematic}.

%
%
%
%

Motor primitives and their generating guidelines are part
of the basic constituents of a cognitive system 
\cite{gros2010cognition}. Here we investigate
whether self-organizing principles may be used also 
on a higher level. As a background we consider a 
setting where an agent has to follow a certain number 
of goals successively, with a typical example being that 
of an animal needing to forage, to watch out for predators, 
to rest and to socialize \cite{sibly1976fitness}.
The agent is hence confronted with tasks that can be 
tackled only sequentially, a problem that may be cast 
into the framework of multi objective optimization 
\cite{deb2014multi}, an approach which is however not 
taken in the present study. We examine instead to 
which extend a self-organized dynamical system may 
solve the time allocation problem implicitly.

\begin{figure}[!t]
\centering
\includegraphics[width=0.75\textwidth]{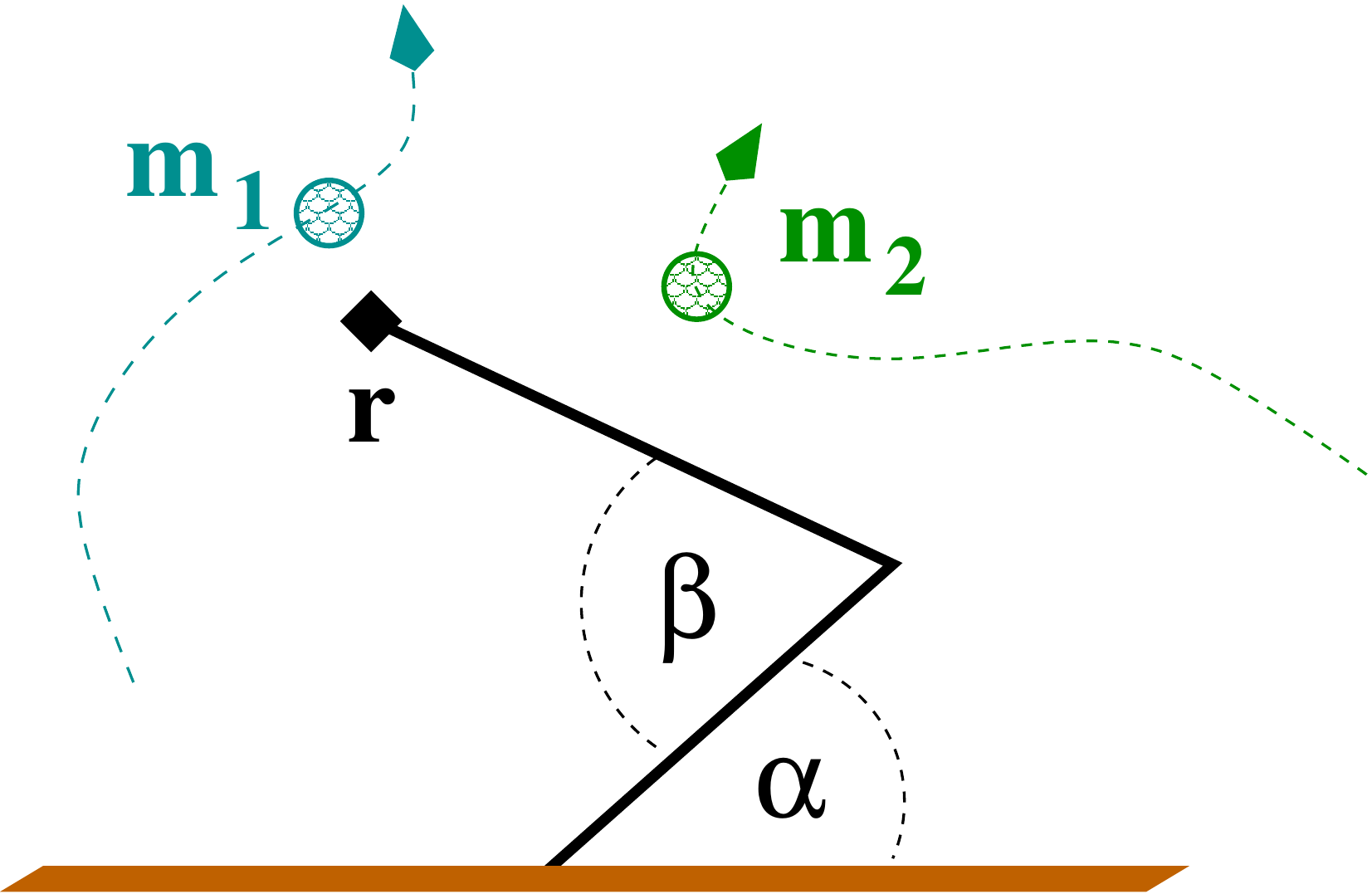}
\caption{{\bf The simulated robot arm.}
The two angles $\alpha$ and $\beta$ are actuated,
with (\ref{dot_alpha_v}) governing the evolution 
of $\alpha$. An equivalent dynamical system is 
in place for $\beta$. The arm has the task to catch one
of the slowly moving objects $\mathbf{m}_i$, to
follow it for a while, with $\mathbf{r}\approx\mathbf{m}_i$, 
and to switch autonomously to a distinct object.
}
\label{fig_arm_sketch}
\end{figure}

As a basic protocol we consider an agent having to 
solve a series of indistinguishable tasks, with the
agent being given by a simulated two-dimensional robot 
arm, as depicted in Fig.~\ref{fig_arm_sketch}. Within 
the reach of the arm there are a number of slowly
moving objects the end actuator needs to reach and follow.
Upon success, the self-organized dynamics of the arm should
become 'bored' of the object, move away and search for a 
new one. We consider this protocol as a proxy for an 
agent showing a non-trivial sequence of behaviors 
generated not by top-down commands, but that emerges from
underlying self-organizing principles.

\section*{Materials and methods}

The simulated robot arm sketched in Fig.~\ref{fig_arm_sketch}
has two degrees of freedom, the angles $\alpha$ and $\beta$,
with the position $\mathbf{r}=(r_1,r_2)$ of the end effector, 
the hand, being given by
\begin{align}
r_1 &= l_1 \cos (\alpha) - l_2 \cos (\beta - \alpha) \\ 
r_2 &= l_1 \sin (\alpha) + l_2 \sin (\beta - \alpha)\,,
\label{r_1_2}
\end{align}
where $l_1$ and $l_2$ are the respective arm lengths.
We define a generalized potential $U$ as
\begin{equation}
U = U_{m}\prod_i T^2\big(R_i\big),\qquad\quad
R_i = \sqrt{\left( \mathbf{r} -\mathbf{m}_{i} \right)^2},
\label{U_R_i}
\end{equation}
where $R_i$ is the Euclidean distance between the position 
$\mathbf{m}_{i}$ of the $i$th target object and 
$\mathbf{r}=\mathbf{r}(\alpha,\beta)$.
In (\ref{U_R_i}) we used a squashing function $T$, 
\begin{equation}
T \left( z \right) = \kappa_z \tanh \left( z / s_z \right), 
\qquad\quad 
\frac{\partial T}{\partial \theta} = 
\frac{\kappa_z}{s_z} 
\left( 1 - \dfrac{T^2}{\kappa^2_z} \right) \frac{\partial z}{\partial \theta}~,
\label{T_squashing}
\end{equation}
which is characterized by a maximal value $\kappa_z$ and a scale
$s_z$. We use $T(z)$ throughout this study for the renormalization
of several dynamical quantities, with the purpose to avoid exceedingly
large forces or velocities. For the case of the distance we select
a maximum value $\kappa_R\to1$, such that we have
$T(R_i)=\tanh(R_i/s_R)$, as entering (\ref{U_R_i}). $U_m$
is then the maximal value for the potential $U=U(\alpha,\beta)$.

\subsection*{Robot arm dynamics}

The dynamics of the angle $\alpha$ is controlled by
\begin{align}
\dot{\alpha}   &=  T \left( v_\alpha \right),
\qquad\quad
\dot{v}_\alpha = f(U) \, T \left( v_\alpha \right) - 
         \nabla_{\alpha} \, U (\alpha, \beta)\,,
\label{dot_alpha_v}
\end{align}
where the objective function $U(\alpha, \beta)$ has the form
of a mechanical potential, with $\nabla_{\alpha}$ denoting the 
gradient with respect to $\alpha$. Equivalent equations govern 
the time evolution of $\beta$. Eq.~(\ref{dot_alpha_v}) corresponds
to a mechanical system with a potential $U$ and a dissipation
function $f(U)$, for which the velocity $v_\alpha$ has been 
renormalized by $T(z)$. 

Mechanical systems with dissipation functions $f(U)$ depending 
exclusively on the potential $U$, as in (\ref{dot_alpha_v}),
can be considered on a general level as versatile prototype dynamical 
systems which exhibit, beside other, complex bifurcation
cascades \cite{sandor2015versatile}.
      Several forms may be selected for the dissipation function $f(U)$, as 
      proposed further below. The system is adaptive \cite{gros2015complex}, 
      dispersing and taking up energy respectively for $f<0$ and $f>0$.
\begin{itemize}
\item In the dissipative stage, when $f(U)<0$, the arm will follow a
      damped trajectory towards the next minimum of the potential $U=U(R)$,
      that is towards the next object $\mathbf{m}_i$.
\item For a dynamical dissipation function $f(U)$, that is for a $f=f(U)$ 
      which depends functionally but not necessarily explicitly on the potential 
      $U$, one can achieve that the state $\mathbf{r}\approx\mathbf{m}_i$ 
      becomes progressively unstable, such that the arm eventually moves 
      away from the object upon taking up energy after $f(U)$ becomes positive.
\end{itemize}
The mechanical potential in (\ref{dot_alpha_v}) treats all targets 
$\mathbf{m}_i$ on an equal footing, the setup studied here.

\subsection*{Dissipation function dynamics}

The generic principle for selecting the dissipation function 
$f(U)$ is that the system needs to be dissipative when 
far away from all objects $\mathbf{m}_i$, with the 
configuration $\mathbf{r}\approx\mathbf{m}_i$ becoming 
unstable once a specific target has been reached and 
followed for a certain time. Distinct ways to implement
this principle are conceivable, here we study
three possibilities.

\begin{itemize}
\item {\bf Exponentially damped (ED).}
   One may presume that the dissipation should become 
   small far away from the objects, viz for large 
   potentials $U$, as expressed by the ansatz
\begin{equation}
f(U) = f_0 \exp (- \mu \, U),\qquad\quad
\tau_f \, \dot{f}_0 = E_t - U\,.
\label{f_U_ED}
\end{equation}
The prefactor $f_0$ changes sign when the potential 
$U$ stays below the reference energy $E_t$ for a period 
comparable to $\tau_f$, viz when the end effector remains 
close to an object. Once $f_0$ turns positive, the arm will 
start to move away from the current object $\mathbf{m}_i$.

\item {\bf Trailing potential (TP).}
  In this setup the dissipation function is explicitly 
  time dependent, with the evolution equation being
  determined by the trailing potential $U_T=U_T(t)$,
\begin{equation}
\tau_f \, \dot{f} = E_t - U_T,\qquad\quad
\tau_{T} \, \dot{U}_T = U - U_T\,,
\label{f_U_TP}
\end{equation}
where the integration time scales are regulated 
by $\tau_f$ and $\tau_T$. The system is dissipative 
when $U_T$ is large, taking up energy once it 
falls below the reference energy $E_t$.

\item {\bf Adapting threshold (AT).}
  One postulates that $f(U)$ becomes positive when the
  potential $U$ falls below a time dependent threshold
  $U_\theta=U_\theta(t)$:
\begin{equation}
f(U) = f_0 \, \left( U_\theta - U \right)\exp(-\mu U),
\qquad\quad
\tau_\theta \, \dot{U}_\theta = E_t - U\,,
\label{f_U_AT}
\end{equation}
where $E_t$ is a reference energy. The overall
scale for $f(U)$ is regulated by $f_0$, with
$\tau_\theta$ determining the time needed for
starting to take up energy, after the target
has been reached dissipatively.

\end{itemize}

Further below we will present comparative
results for the above three types of dissipation
function dynamics, with in-detail investigations
of robustness and other dynamical properties
concentrating on ED.

\begin{figure}[!t]
\centering
\includegraphics[width=.8\textwidth]{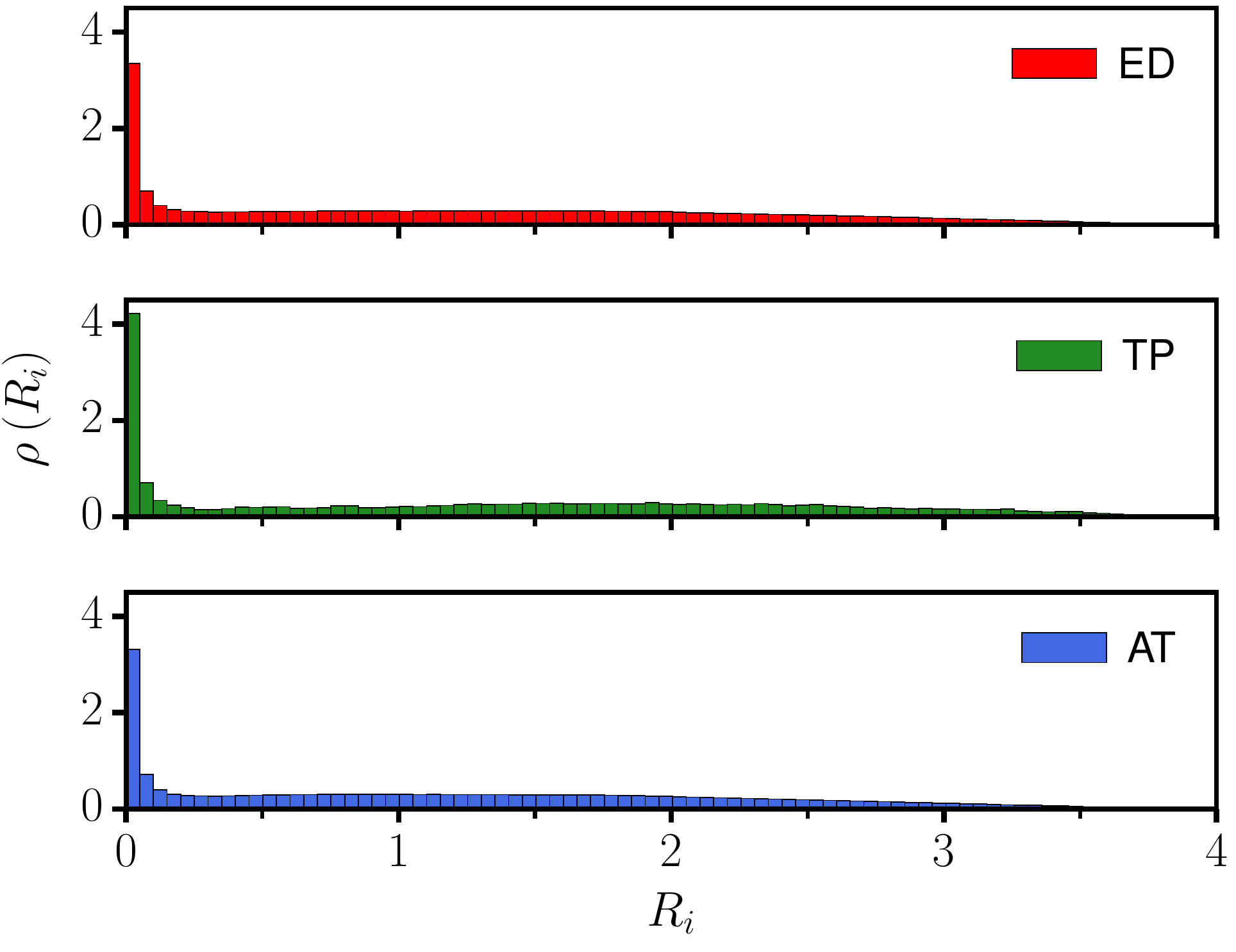}
\caption{{\bf Distance statistics.}
The probability distribution $\rho(R_i)$ for he distance 
$R_i$ between the end effector and a selected object $i$,
as averaged over time. The targets are indistinguishable, 
which implies that $\rho(R_i)=\rho(R_j)$ for all $i,j\in[1,n]$, 
where $n=3$ is the number of moving objects. Shown are the results 
for three different dissipation functions dynamics, ED 
(top,
\href{https://doi.org/10.6084/m9.figshare.7706630.v2}
{click for animation} to see \nameref{S1_Video}),
TP (middle, 
\href{https://doi.org/10.6084/m9.figshare.7706624.v1}
{click for animation} to see \nameref{S2_Video}),
and AT (bottom, 
\href{https://doi.org/10.6084/m9.figshare.7706627.v1}
{click for animation} to see \nameref{S3_Video}),
as defined respectively by (\ref{f_U_ED}),
(\ref{f_U_TP}) and (\ref{f_U_AT}).
The parameters are listed in Table.~\ref{tableParameters}.
}
\label{fig_distance_probabilities}
\end{figure}

\subsection*{Moving objects}

For the dynamics of the moving objects, the robot arm 
has to grab, we used two closely related algorithms.

\begin{itemize}
\item {\bf Polar representation of the velocity (M-PV).} 
      In the first case the absolute velocity $|v_i|$ of 
      an object $\mathbf{m}_i$ is drawn from an uniform 
      distribution in $[0,a]$, with the angle $\varphi_i$ 
      being drawn from $[0,2\pi]$. 

\item {\bf Cartesian representation of the velocity (M-CV).}
      In the second approach the Cartesian xy-components of 
      $\mathbf{v}_i$ are drawn independently from an uniform 
      distribution in $[-b, b]$. 

\end{itemize}
The resulting velocity $\mathbf{v}_i$ is applied in both cases
for a time span $t_i$ which is drawn uniformly from $[0, t_{max}]$.
The diffusion of the object is restricted in addition to a circular 
area of radius $r_{area}$, reflecting at the boundary. We generally 
selected $r_{area}$ to coincide with the reach of the robot arm.
For the other parameters we took $a=b=0.001$ and $t_{max}=10$.

As the simulation results for M-PV and M-CV are very similar,
we show in the following the ones for M-PV.

\begin{table}[!b]
\caption{{\bf Simulation parameters.} The parameters
$\kappa_v$ and $s_v$ entering the renormalization
of the velocity of the mechanical system (\ref{dot_alpha_v}) 
have been adapted slightly for the three different dissipation
function dynamics, ED, TP and AT. Listed are furthermore
all parameters entering the respective defining equations
(\ref{f_U_ED}), (\ref{f_U_TP}) and (\ref{f_U_AT}). Note
that $\mu$ is given in units of $1/U_m$.}
\smallskip
\centering
\begin{tabular}{l|lc|ccccc}
   & $\kappa_v$ & $s_v$ & $\mu U_m$ & $\tau_f$ & 
     $\tau_T$   & $\tau_\theta$ & $f_0$ \\
\hline
{\bf ED} & 2.8 & 1 & 25        & 1.2       & $\bullet$ & $\bullet$ & $\bullet$\\
{\bf TP} & 4.3 & 3 & $\bullet$ & 6.0       & 4.0       & $\bullet$ & $\bullet$\\
{\bf AT} & 4.0 & 2 & 34        & $\bullet$ & $\bullet$ & 1      & 0.5\\
\end{tabular}
\label{tableParameters}
\end{table}

\begin{figure}[!t]
\centering
\includegraphics[width=.8\textwidth]{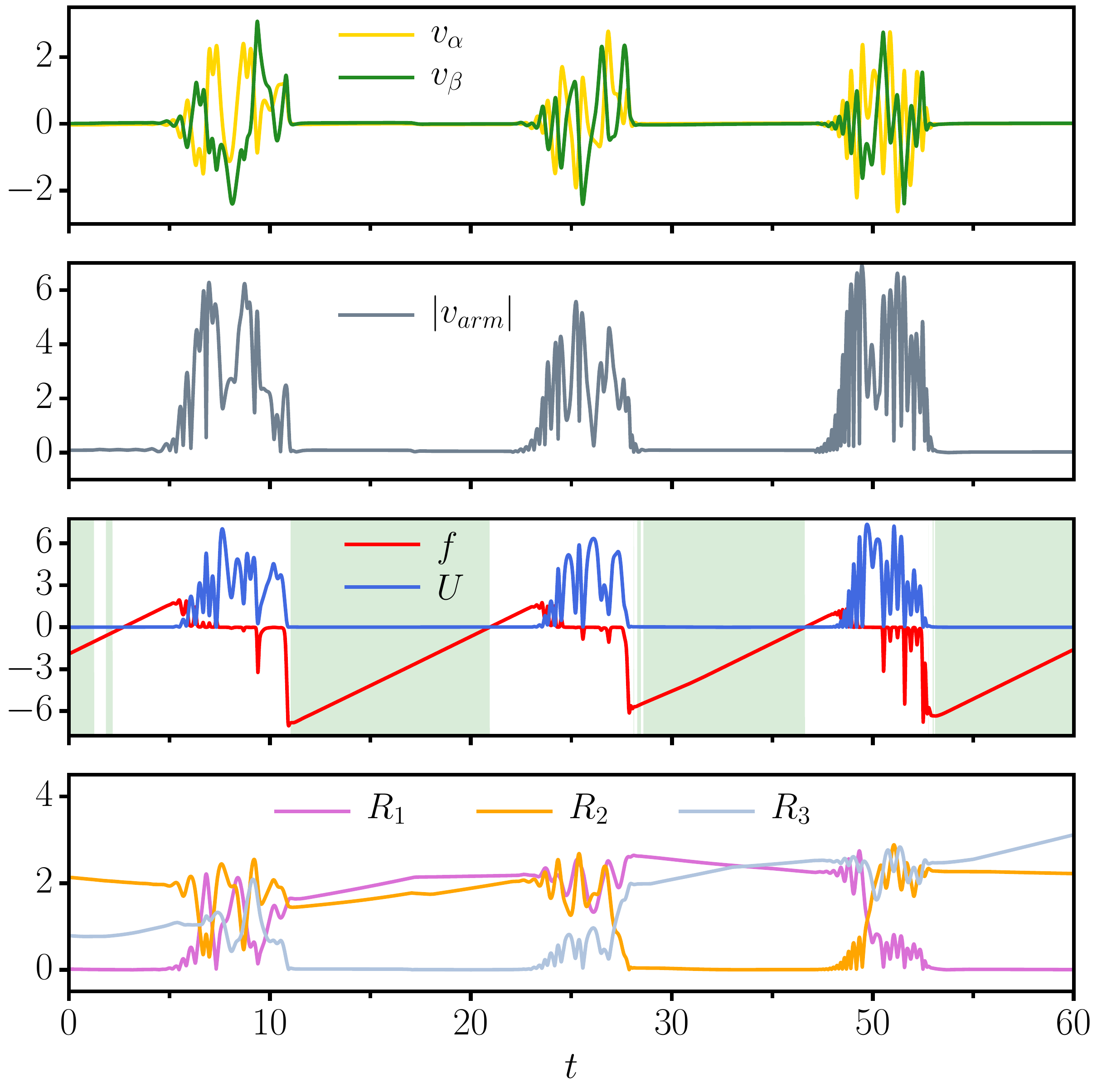}
\caption{{\bf Time series for three moving objects.}
As a function of simulation time~$t$, the evolution of
key variables for the ED dissipation-function dynamics,
compare (\ref{f_U_ED}). 
(top) The angular velocities $v_\alpha$ and $v_\beta$.
(second from top) The modulus $|v_{arm}|$ of velocity 
$v_{arm}$ of the end effector.
(second from bottom) The dissipation function $f$ and 
the potential $U$, see (\ref{U_R_i}), with the shading 
indicating that the criterion (\ref{following_criterion}) is fulfilled.
The separation of time scales characterizing the dynamics 
of $f$, for which a fast drop to negative values is 
followed by a slow recovery, drives the distinction 
between irregular searching phases and the laminar 
flow observed when the end-effector is close to a 
specific target.
(bottom) The  distances $R_i$ to the $n=3$ moving objects.
}
\label{fig_timeline_ED_normal}
\end{figure}

\subsection*{Parameters}

The overall length $L=l_1+l_2$ of the arm is set to $L=2$,
with the lengths of the two segments being identical,
$l_1 = l_2 = 1$. The parameters for the squashing
function (\ref{T_squashing}) for the distance are
$\kappa_R = 1$ and $s_R = \sqrt{3/n}\,L/2$. For
$n=3$ moving objects we have hence $s_R=L/2=1$.

For the maximum of the potential $U_m$ and for the reference
energy $E_t$ we used $U_{m}=17$ and $E_t=0.05U_m$, respectively,
with all other parameters being listed in Table \ref{tableParameters}.
For the simulation a time step of $dt=0.01$ has been used.

\section*{Results}

For the parameters given in Table~\ref{tableParameters}
we find transients in which the arm tends to stay close 
to a target it has approached. The flow in phase space is
laminar when the arm is close to a target, accelerating
however considerably once the dissipation function 
$f(U)$ turns positive, compare (\ref{dot_alpha_v})
together with (\ref{f_U_ED}), (\ref{f_U_TP}) and (\ref{f_U_AT}).
For a first understanding we present in Fig.~\ref{fig_distance_probabilities}
the probability $\rho(R_i)$ to observe the distance $R_i$
between the end effector and a given target $i$, see
(\ref{U_R_i}). With all $n=3$ targets being equivalent,
one has $\rho(R_i)=\rho(R_j)$, for all $i,j\in[1,n]$.

\subsection*{Following vs.\ explorative phase}

The distribution of the distance $R_i$ presented in
Fig.~\ref{fig_distance_probabilities} shows that the motion
of the arm can be subdivided into a phase of small $R_i$
and a phase of medium to large distances of all sizes,
modulo fine details. That this is the case for three 
different types of dissipation function dynamics proves 
that the underlying generating principles is both robust
and versatile. For the three variants considered here, 
(\ref{f_U_ED}), (\ref{f_U_TP}) and (\ref{f_U_AT}),
the arm will start to take up energy whenever it did hover
for a certain time close to a target, dissipating on the other 
side energy when far away.

The evolution of key variables as a function of simulation
time is presented in Fig.~\ref{fig_timeline_ED_normal}. Shown are,
for the ED dissipation function dynamics, the velocities
$v_\alpha$, $v_\beta$ and $v_{arm}$, of the actuators
and respectively of the arm, together with the evolution
of the dissipation function $f$, of the potential $U$,
and of the distances $R_i$ between the hand of the arm
and the individual objects.

One can distinguish in Fig.~\ref{fig_timeline_ED_normal} 
laminar `following phases' and highly irregular 
`explorative phases'. Particularly
evident is the driving role of the dissipation 
function, which remains negative for most of the smooth
following phase. Visible is also a certain time lag
between the crossing of $f$ from negative to
positive values, which results from the time the system
needs to take up enough energy for the angular velocities
$v_\alpha$ and $v_\beta$, and the potential $U$ to become 
visible.

\begin{figure}[!t]
\centering
\includegraphics[width=1.0\textwidth]{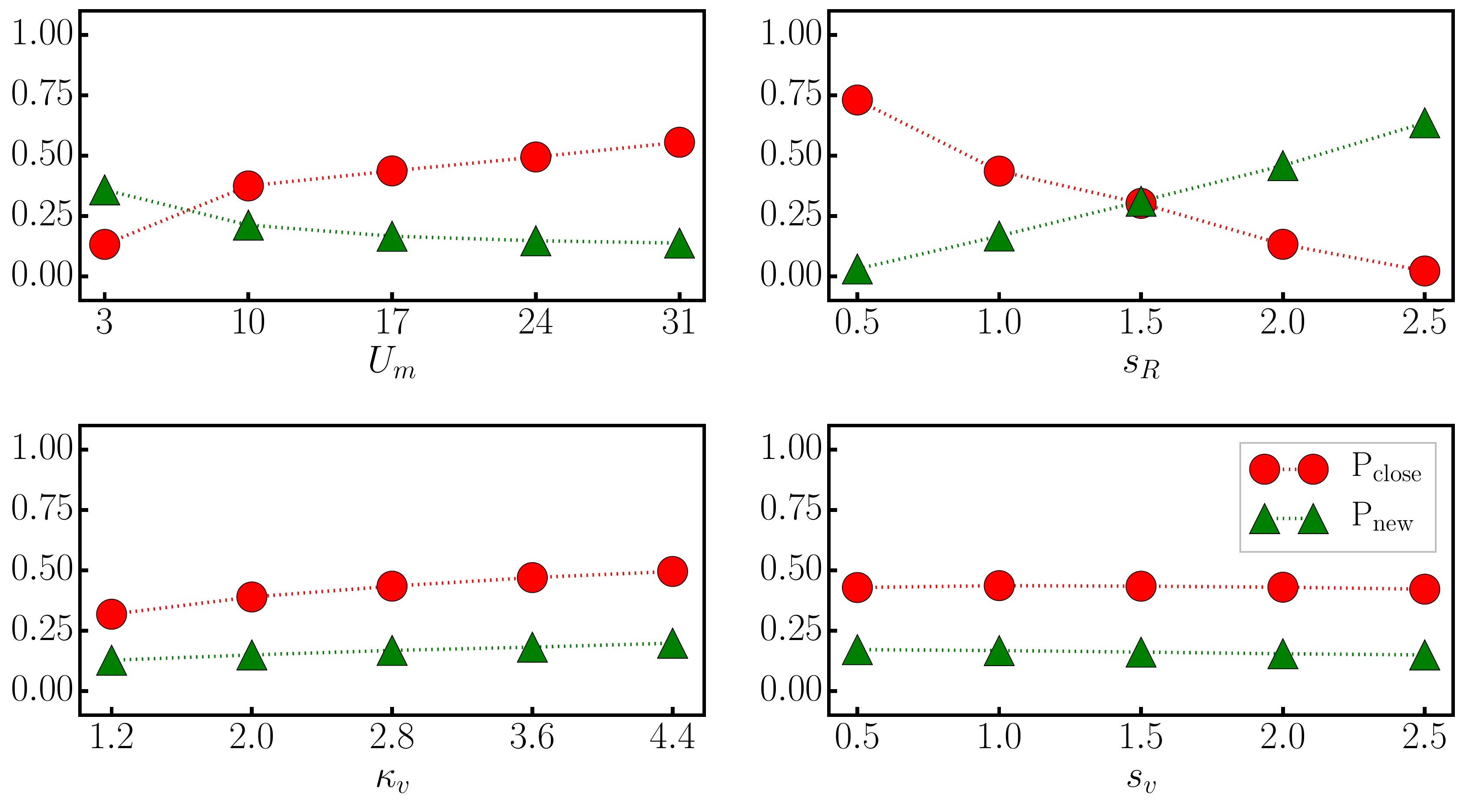}
\caption{{\bf Parameter sweep.}
For the ED dissipation function dynamics, the probability 
$P_{close}$ for the arm to be close to one of the $n=3$ targets 
(red circles), as defined by (\ref{following_criterion}), and 
$P_{new}$, which measures the chance that two targets approached 
one after the another are different (green triangles). With respect 
to the reference values $U_m=17$, $s_R=L/2=1$, $\kappa_v=2.8$ 
and $s_v=1$, the values of the parameters have been changed 
individually.
}
\label{fig_parSweep_Um_sR_kv_sv}
\end{figure}

\subsection*{Robustness with respect to parameter changes}

For a criterion that determines whether the end 
effector follows a given target we use
\begin{equation}
U<E_t,\qquad\quad 
f(U)<0, \qquad\quad
|v_{arm}|<v_{tar}^{max}~, 
\label{following_criterion}
\end{equation}
which demands that the potential $U$ is small with respect 
to the threshold energy $E_t$ and that the system is
momentarily dissipative, viz that the dissipation function $f(U)$
is negative. The last term in (\ref{following_criterion})
rules out coincidental crossings at high velocities,
which occur when magnitude of the velocity $v_{arm}$ of 
the end effector is larger than the maximal velocity 
$v_{tar}^{max}$ of the targets. With the dynamics of the 
targets being generated, as described, $v_{tar}^{max}$ 
is known. For practical applications it would be in any 
case sufficient to use an empirical estimate for $v_{tar}^{max}$.

Using the criterion (\ref{following_criterion}), one can
define a probability $P_{close}$ that measures the relative
fraction of time the arm follows a target, with following
and the exploration being the two dominant states of the 
system, as evident from Fig.~\ref{fig_timeline_ED_normal}.

In Fig.~\ref{fig_parSweep_Um_sR_kv_sv} we present for
the ED dissipation function dynamics the numerical 
result for $P_{close}$. Starting from the reference
set of parameters $U_m=17$, $s_R=L/2=1$, $\kappa_v=2.8$
and $s_v=1$, compare also Table~\ref{tableParameters},
the parameters have been modified one by one and 
the probability for the arm to follow a target evaluated. 
Also included in Fig.~\ref{fig_parSweep_Um_sR_kv_sv} is
the probability $P_{new}$, namely that two targets approached
successively differ.

\begin{figure}[!t]
\centering
\includegraphics[width=1.0\textwidth]{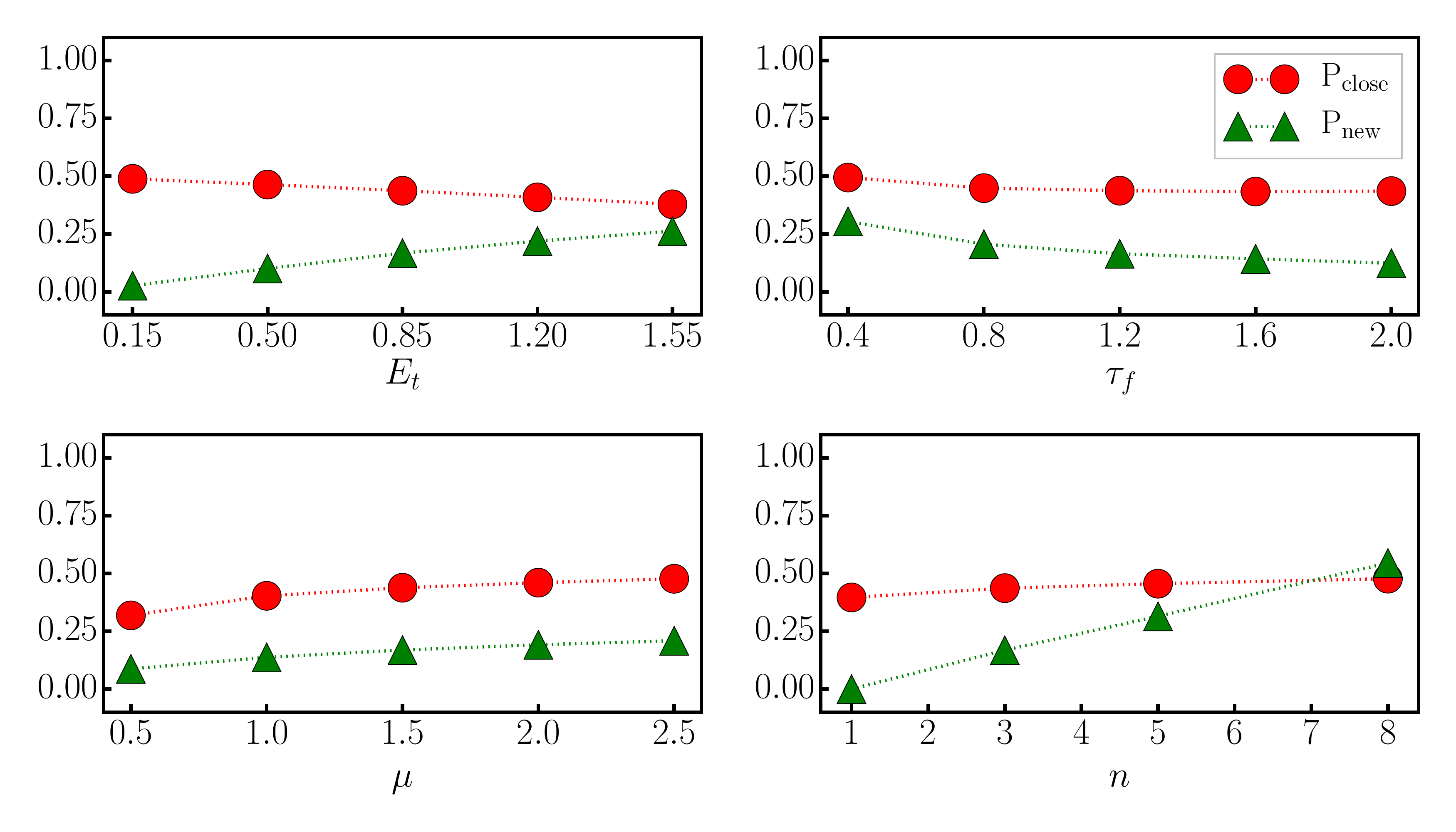}
\caption{{\bf Robustness of the dissipation function dynamics.}
For the ED dissipation function dynamics, the probability 
$P_{close}$ for the arm to be close to a target (red circles), 
as defined by (\ref{following_criterion}), and $P_{new}$, which 
measures the chance that two targets approached one after 
the another are different (green triangles). With respect to 
the reference values $E_t=0.05U_m=0.85$, $\mu=25/U_m=1.47$
$\tau_f=1.2$, the values of the parameters have been changed 
individually for $n=3$. Also included are the values
of $P_{close}$ and $P_{new}$ upon changing the number $n$
of targets. Here $s_R = \sqrt{3/n}L/2$.
}
\label{fig_parSweep_Et_tf_mu_n}
\end{figure}

\begin{itemize}
\item The probability $P_{close}$ for the arm to be 
      in the following phase increases monotonically with
      the strength $U_m$ of the potential, an intuitive
      result. $P_{new}$ decreases conversely, with the reason
      being that a larger $U_m$ makes it more difficult to 
      escape the local potential well.
\item Increasing the characteristic length $s_R$ for 
      the distance between the arm and a target, 
      which enters the squashing function (\ref{T_squashing}),
      decreases $P_{close}$ dramatically. This is because
      the local potential wells attracting the end actuator 
      to a target in first place tend to disappear for large 
      $s_R$. $P_{new}$ increases on the other side.
\item The squashing parameters $\kappa_v$ and $s_v$ for 
      the velocity of the actuators can be changed considerable
      without affecting either $P_{close}$ or $P_{new}$,
      implying that the system is robust with respect to
      both $\kappa_v$ and $s_v$.
\end{itemize}

The data shown in Fig.~\ref{fig_parSweep_Um_sR_kv_sv}
describes the influence of global parameters. In
Fig.~\ref{fig_parSweep_Et_tf_mu_n} we present for
completeness the effect of changing the parameters
$E_t$, $\mu$ and $\tau_f$ of the ED dissipation 
function dynamics, see (\ref{f_U_ED}). We find
the generating principle to be robust, viz that
the dependency of $P_{close}$ and $P_{new}$ on
$E_t$, $\mu$ and $\tau_f$ is moderate.

Also included in Fig.~\ref{fig_parSweep_Et_tf_mu_n} are
the values of $P_{close}$ and $P_{new}$ obtained upon 
changing the number $n$ of targets. One observes that the
relative fraction of time $P_{close}$ the arm spends close
to a target remains flat. For $n=1$ the probability to
change targets vanishes, as it must, becoming on the
other side substantial for large numbers of targets
$n$.

The here presented sequential task-switching behavior, 
generated by the prototype dynamical system (\ref{dot_alpha_v})
does not rely on the particular choice of the generalized 
dissipation function dynamics. As demonstrated by 
Fig.~\ref{fig_distance_probabilities}, similar distance 
distributions $\rho(R_i)$ may result from very 
different dissipation function implementations.
This is also reflected by the fraction of time spent 
with following and the probability of switching targets, 
$P_{close} = 0.44/0.69/0.44$ and $P_{new} = 0.17/0.07/0.14$,
when comparing the dissipation functions
ED/TP/AT see Eqs.~(\ref{f_U_ED}), (\ref{f_U_TP}) and 
(\ref{f_U_AT}) respectively, for the parameters given in 
Table~\ref{tableParameters}.

\subsection*{Robustness with respect to target properties}

\begin{figure}[!t]
\centering
\includegraphics[width=.8\textwidth]{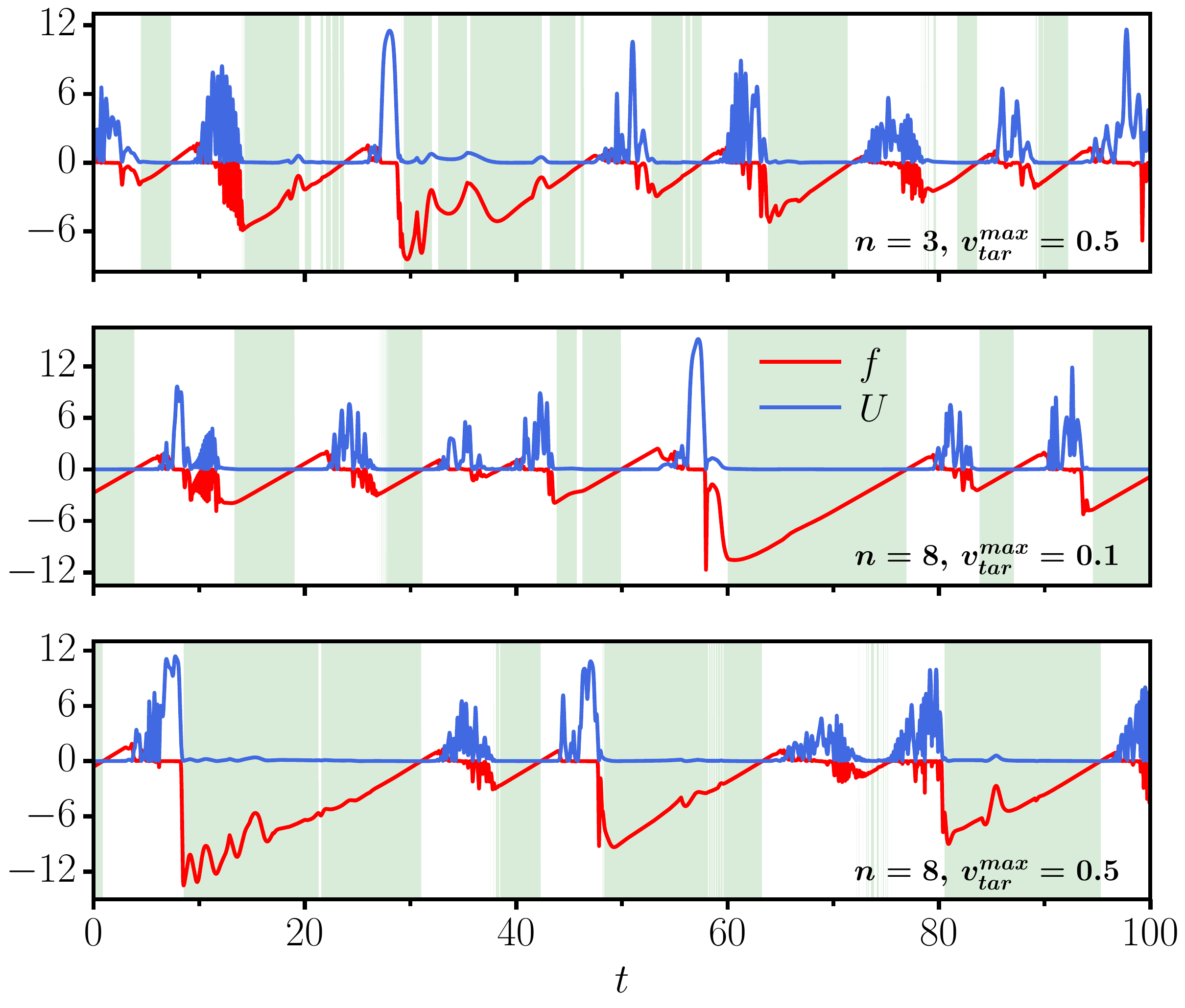}
\caption{{\bf Variable object characteristics.}
As a function of simulation time~$t$, the evolution of
the dissipation function $f$ (red) and of the potential 
$U$ (blue) for the ED dissipation-function dynamics.
The shaded regions indicate that the criterion
(\ref{following_criterion}) for the arm to be in
the following phase is fulfilled.
(top) For $n=3$ objects for which the maximal velocity is $0.5$, viz
five times larger than in Fig.~\ref{fig_timeline_ED_normal}
(\href{https://doi.org/10.6084/m9.figshare.7706618.v2}
{click for animation} to see \nameref{S4_Video}).
(middle) For $n=8$ objects with a maximal velocity $0.1$
(\href{https://doi.org/10.6084/m9.figshare.7706621.v2}
{click for animation} to see \nameref{S5_Video}).
(bottom) For $n=8$ objects with a maximal velocity $0.5$
(\href{https://doi.org/10.6084/m9.figshare.7706633.v2}
{click for animation} to see \nameref{S6_Video}).
}
\label{fig_timeline_ED_special}
\end{figure}

It is clear that the arm would not be able to follow 
a target if the maximal velocity $v_{tar}^{max}$ is 
too large. We find, however, that the here proposed
generating principle works for a substantial range of
$v_{tar}^{max}$. For the ED dissipation function
dynamics we present in Fig.~\ref{fig_timeline_ED_special}
the time series of the dissipation function and of the
potential  both for the case of $v_{tar}^{max}=0.1$,
as used hitherto, and for $v_{tar}^{max}=0.5$.
We find that only details of the overall dynamics
change. This holds also when increasing
the number of moving objects from $n=3$ to $n=8$.

\subsection*{A single non-moving target}

From the dynamical system perspective it is of
interest to investigate the case of a single 
stationary target. With noise being absent,
the system is deterministic. 

\begin{figure}[!t]
\centering
\includegraphics[width=.8\textwidth]{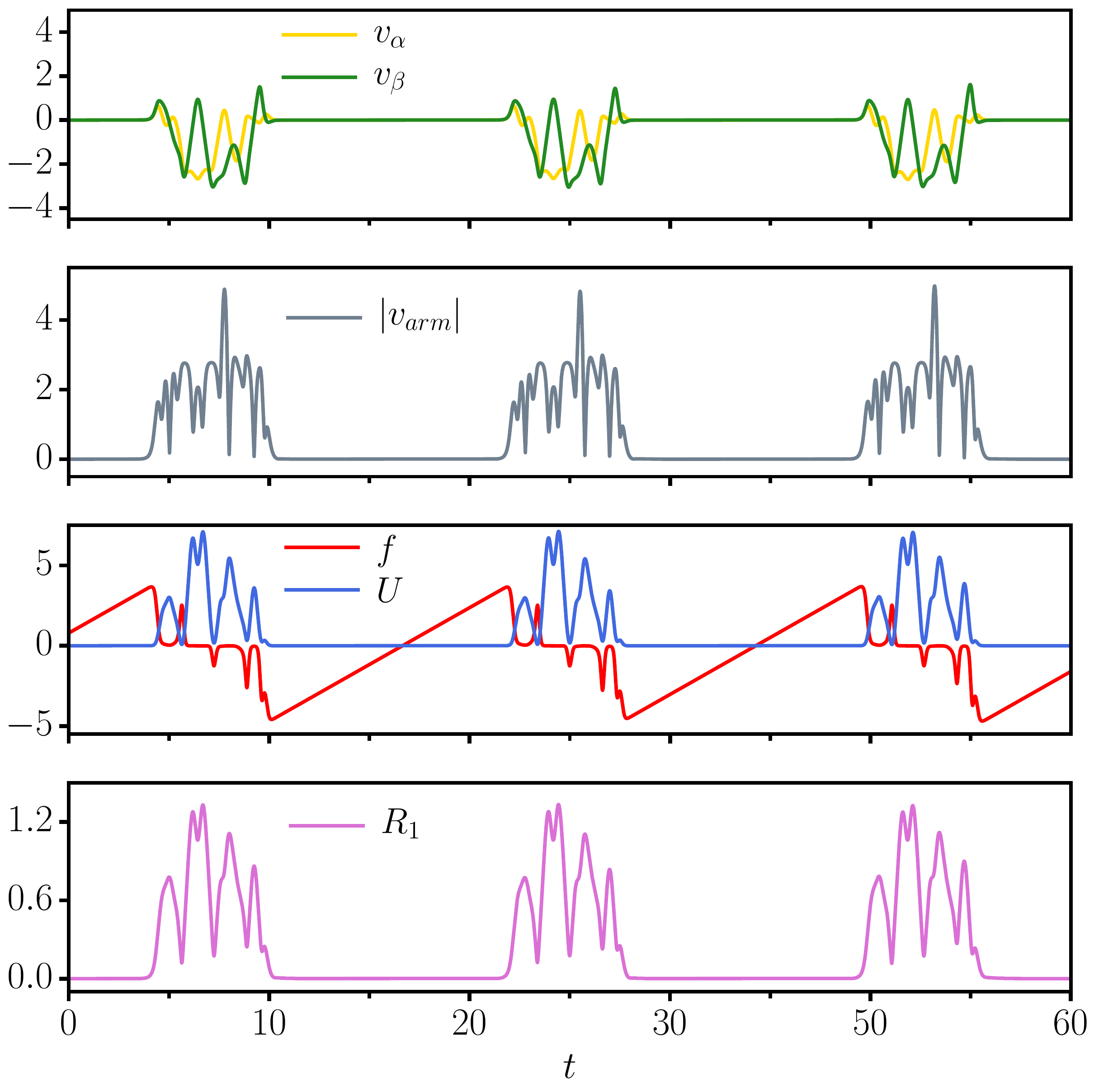}
\caption{{\bf A single non-moving object.}
As a function of simulation time~$t$, the evolution of
key variables for the ED dissipation-function dynamics,
as for Fig.~\ref{fig_timeline_ED_normal}, but here for
a single non-moving object located at $(3/8,3/8)\,L$. 
The system is fully deterministic, with the robot arm 
settling into a limit cycle.  The criterion 
(\ref{following_criterion}) for the arm to
be close to the object is not applicable, as
$v_{tar}^{max}=0$.
}
\label{fig_timeline_ED_fixed}
\end{figure}

\begin{itemize}
\item \textbf{Fixpoints.} In case of a purely 
dissipative dynamics, with $f(U)=f_0<0$, the system disposes 
of two stable fixpoints, defined by vanshing angular
velocities $v_{\alpha},v_\beta\to0$, that correspond
to a right- and respectively to a left bend.


\item \textbf{Limit cycle attractors.} With the dynamical
dissipation function ED, it is evident that the
robot arm settles into a limit cycle in which the 
destabilized fixpoints are revisited, see
Fig.~\ref{fig_timeline_ED_fixed}. There exist, 
hence, multiple symmetry related limit cycles even 
for a single resting target (only one of them is shown).
\end{itemize}

Therefore, in the presence of multiple fixed targets, several
different activity sequences may be generated, even for
the same starting position~$\mathbf{r}(0)$ of the arm, 
viz for different initial conditions of the internal variables.


\section*{Discussion}

Action switching in embodied agents may be guided
by fitness considerations, f.i.\ when the task is
to collect a series of different food sources 
\cite{agmon2014evolution}. Typically, the action 
selected at a given time will be then the one
with the most pressing need. We have followed here
a different approach, examining an overarching generation 
principle and not the generation of action sequences 
driven by an utility optimization that is local in time.

\subsection*{The stationarity principle}

The question how to decide in which action to 
engage has been termed the motivational problem 
\cite{gros2012emotional}. The utility of many 
activities, like foraging, socializing and resting, 
that are regularly repeated, address distinct
needs, which implies that they cannot be lumped
together into an overarching utility function.
In terms of multi-objective optimization \cite{deb2014multi}
the agent must dedicate time to a range of 
activities, with the constraint that the resulting 
distribution of utilities remains within a given
range. This constraint may be expressed as a 
stationarity principle, namely that the statistical
properties of the time series of activities should
become stationary for extended time spans.

The result presented here for the self-organized
robot arm can be viewed as an implementation of
the stationarity principle. With the dynamics being
irregular, viz chaotic, in the explorative phase,
the exact sequence of objects followed is not
pre-determined. The long term statistics, such
as the distance distribution presented in 
Fig.~\ref{fig_distance_probabilities}, is
however stationary.

The stationarity principle is a guiding principle
that can be used in various settings. Statistical
learning, e.g.\ of receptive fields \cite{brito2016nonlinear}, 
is characterized by statistically stationary sensory inputs, 
with learning continuing until the statistics of the 
output activity becomes also stationary \cite{echeveste2015fisher}.
It has been shown, that one can use the Fisher information
of the neural firing rate to encode the stationarity
principle \cite{echeveste2014generating} and that one
obtains Hebbian learning when minimizing the Fisher information, 
viz when the stationarity condition is enforced.

\subsection*{Transient-state dynamics}

With the agent being formulated in term of a mechanical 
system, see Eq.~(\ref{dot_alpha_v}), one can abstract from 
the behavioral level and describe the robot arm within 
dynamical system theory \cite{gros2015complex}. The 
striking alternation of dynamical states, as visible in 
Fig.~\ref{fig_timeline_ED_normal}, can be interpreted in this 
context as an example of transient-state dynamics \cite{gros2007neural}.
The following phase corresponds on a dynamical level
to a transient attractor that becomes unstable on an
extended time scale, namely when the dissipation function
turns positive. 

The here discussed mechanism, the coupling of an attracting
state to a slow variable, is the core route for generating
transient-state dynamics in general \cite{gros2009cognitive},
with the flow being laminar during the transient dynamics,
and irregular during the transition periods.
We note that transient-state dynamics may be viewed as 
a form of metastability, which may arise either from the
brain dynamics as such \cite{kelso2012multistability},
or from sensorimotor couplings in response to tasks demanding 
behavioral flexibility \cite{aguilera2016extended}.




\subsection*{Distinguishable vs.\ non-distinguishable targets}

It would be possible to introduce a bias $b_i=b_i(t)$ that 
allows to differentiate between distinct objects. In this 
case one would work with the generalized Euclidean distance
\begin{equation}
R_i \ \to\ \sqrt{R_i^2+b_i^2}\,.
\label{RR_cc}
\end{equation}
instead of (\ref{U_R_i}),
for which the bias $b_i$ encodes the depth of the potential, and with 
this indirectly also the relative importance of the respective 
object. For an appropriate evolution equation for $b_i(t)$, the
respective target would become repelling once the end effector
of the robot has reached it. Two routes on how the dynamical
system (\ref{dot_alpha_v}) induces an autonomously 
generated sequence of behaviors are hence possible. 
\begin{itemize}
\item {\bf Distinguishable targets.} One works with a  constant
     dissipation function, $f(U)\to f_0$, with every object
     being characterized by a time-dependent attribute, namely
     $b_i=b_i(t)$.
\item {\bf Indistinguishable targets.} When all $b_i\equiv0$
     there is no variable distinguishing the individual objects.
     The sequence of behaviors is then a consequence of
     dynamical instabilities resulting from the dynamics of 
     the dissipation function. 
\end{itemize}
In this study we concentrate on the second case as the basic generative
mechanism, noting that the resulting residence times, viz when 
$\mathbf{r}\approx\mathbf{m}_i$, could be fine-tuned in a 
second step by allowing the $b_i$ to be weakly time dependent.
This protocol is left for future studies.

\section*{Conclusion}

One of the biggest challenges in the design of controllers
for autonomous agents is the combination of different 
goal oriented behaviors into a series of self-organized 
activities \cite{bekey2005autonomous}.
Here, we investigated how such a higher order controller 
may be constructed within a dynamical systems framework, 
by adapting a recently introduced versatile prototype 
system \cite{sandor2015versatile} to the problem of 
an object-following arm. By introducing a model with 
a dynamically changing generalized dissipation function 
we provide a proof of concept demonstration 
of how target following can be turned into a sequential 
task switching behavior in terms of transient-state 
dynamics \cite{gros2007neural}. 

Within this framework the goal oriented activities 
are represented by a target-following behavior of 
a simulated arm, while the switching dynamics between targets 
corresponds to an explorative phase upon getting bored 
of the respective task. 

Such a self-organized behavior can be generated 
both at the level of motion primitives, in case of robotic 
locomotion \cite{sandor2015sensorimotor}, and on the
level of action selection \cite{agmon2014evolution}, 
as demonstrated here. The resulting behavior is robust 
within a wide range of parameters, as it does not require 
precise fine tuning, which simplifies the selection of 
an adequate parameter set with, e.g., machine learning 
techniques. Being based on self-organized attractors in 
the overarching phase space of agent and environment, 
the sensorimotor loop, our approach is resistant to 
external noise, retaining at the same time the
flexibility to adapt to the environment or to interact 
with other agents \cite{martin2016closed}.

The proposed framework can be generalized 
to produce series of activities with a well-defined 
order or a given multi-modal probability distribution
by modulating the Euclidean distance as a function 
of the actual importance of the respective task -- 
a research direction left for future studies.

\section*{Acknowledgments}

None.

\nolinenumbers


\begin{thebibliography}{10}

\bibitem{martius2013information}
Martius G, Der R, Ay N.
\newblock Information driven self-organization of complex robotic behaviors.
\newblock PloS one. 2013;8(5):e63400.

\bibitem{ijspeert2014biorobotics}
Ijspeert AJ.
\newblock Biorobotics: Using robots to emulate and investigate agile
  locomotion.
\newblock science. 2014;346(6206):196--203.

\bibitem{aguilera2015self}
Aguilera M, Barandiaran XE, Bedia MG, Seron F.
\newblock Self-organized criticality, plasticity and sensorimotor coupling.
  Explorations with a neurorobotic model in a behavioural preference task.
\newblock PloS one. 2015;10(2):e0117465.

\bibitem{tani2003self}
Tani J, Ito M.
\newblock Self-organization of behavioral primitives as multiple attractor
  dynamics: A robot experiment.
\newblock IEEE Transactions on Systems, Man, and Cybernetics-Part A: Systems
  and Humans. 2003;33(4):481--488.

\bibitem{beer2015information}
Beer RD, Williams PL.
\newblock Information processing and dynamics in minimally cognitive agents.
\newblock Cognitive science. 2015;39(1):1--38.

\bibitem{arshavsky2016central}
Arshavsky Y, Deliagina T, Orlovsky G.
\newblock Central Pattern Generators: Mechanisms of Operation and Their Role in
  Controlling Automatic Movements.
\newblock Neuroscience and Behavioral Physiology. 2016;46(6):696--718.

\bibitem{marder2001central}
Marder E, Bucher D.
\newblock Central pattern generators and the control of rhythmic movements.
\newblock Current biology. 2001;11(23):R986--R996.

\bibitem{ijspeert2008central}
Ijspeert AJ.
\newblock {Central pattern generators for locomotion control in animals and
  robots: A review}.
\newblock Neural Networks. 2008;21(4):642--653.

\bibitem{minassian2017human}
Minassian K, Hofstoetter US, Dzeladini F, Guertin PA, Ijspeert A.
\newblock The human central pattern generator for locomotion: Does it exist and
  contribute to walking?
\newblock The Neuroscientist. 2017;23(6):649--663.

\bibitem{sandor2015sensorimotor}
S{\'a}ndor B, Jahn T, Martin L, Gros C.
\newblock The sensorimotor loop as a dynamical system: how regular motion
  primitives may emerge from self-organized limit cycles.
\newblock Frontiers in Robotics and AI. 2015;2:31.

\bibitem{pfeifer2007self}
Pfeifer R, Lungarella M, Iida F.
\newblock Self-organization, embodiment, and biologically inspired robotics.
\newblock science. 2007;318(5853):1088--1093.

\bibitem{aguilar2016review}
Aguilar J, Zhang T, Qian F, Kingsbury M, McInroe B, Mazouchova N, et~al.
\newblock A review on locomotion robophysics: the study of movement at the
  intersection of robotics, soft matter and dynamical systems.
\newblock Reports on Progress in Physics. 2016;79(11):110001.

\bibitem{owaki2013simple}
Owaki D, Kano T, Nagasawa K, Tero A, Ishiguro A.
\newblock Simple robot suggests physical interlimb communication is essential
  for quadruped walking.
\newblock Journal of The Royal Society Interface. 2013;10(78):20120669.

\bibitem{owaki2017quadruped}
Owaki D, Ishiguro A.
\newblock A quadruped robot exhibiting spontaneous gait transitions from
  walking to trotting to galloping.
\newblock Scientific reports. 2017;7(1):277.

\bibitem{martin2016closed}
Martin L, S{\'a}ndor B, Gros C.
\newblock Closed-loop robots driven by short-term synaptic plasticity: Emergent
  explorative vs. limit-cycle locomotion.
\newblock Frontiers in neurorobotics. 2016;10:12.

\bibitem{sandor2018kick}
S{\'a}ndor B, Nowak M, Koglin T, Martin L, Gros C.
\newblock Kick control: using the attracting states arising within the
  sensorimotor loop of self-organized robots as motor primitives.
\newblock Frontiers in neurorobotics. 2018;12.

\bibitem{muller2017morphological}
M{\"u}ller VC, Hoffmann M.
\newblock What is morphological computation? On how the body contributes to
  cognition and control.
\newblock Artificial life. 2017;23(1):1--24.

\bibitem{ghazi2017morphological}
Ghazi-Zahedi K, Langer C, Ay N.
\newblock Morphological computation: Synergy of body and brain.
\newblock Entropy. 2017;19(9):456.

\bibitem{sprowitz2013towards}
Spr{\"{o}}witz A, Tuleu A, Vespignani M, Ajallooeian M, Badri E, Ijspeert AJ.
\newblock {Towards dynamic trot gait locomotion: Design, control, and
  experiments with Cheetah-cub, a compliant quadruped robot}.
\newblock The International Journal of Robotics Research. 2013;32(8):932--950.

\bibitem{travers2016dynamical}
Travers M, Ansari A, Choset H.
\newblock {A dynamical systems approach to obstacle navigation for a
  series-elastic hexapod robot}.
\newblock In: Decision and Control (CDC), 2016 IEEE 55th Conference on. IEEE;
  2016. p. 5152--5157.

\bibitem{pinneri2018systematic}
Pinneri C, Martius G.
\newblock {Systematic self-exploration of behaviors for robots in a dynamical
  systems framework}.
\newblock In: The 2018 Conference on Artificial Life. Cambridge, MA: MIT Press;
  2018. p. 319--326.

\bibitem{gros2010cognition}
Gros C.
\newblock Cognition and emotion: perspectives of a closing gap.
\newblock Cognitive Computation. 2010;2(2):78--85.

\bibitem{sibly1976fitness}
Sibly R, McFarland D.
\newblock On the fitness of behavior sequences.
\newblock The American Naturalist. 1976;110(974):601--617.

\bibitem{deb2014multi}
Deb K.
\newblock Multi-objective optimization.
\newblock In: Search methodologies. Springer; 2014. p. 403--449.

\bibitem{sandor2015versatile}
S{\'a}ndor B, Gros C.
\newblock A versatile class of prototype dynamical systems for complex
  bifurcation cascades of limit cycles.
\newblock Scientific reports. 2015;5:12316.

\bibitem{gros2015complex}
Gros C.
\newblock Complex and adaptive dynamical systems: A primer.
\newblock Springer; 2015.

\bibitem{agmon2014evolution}
Agmon E, Beer RD.
\newblock The evolution and analysis of action switching in embodied agents.
\newblock Adaptive Behavior. 2014;22(1):3--20.

\bibitem{gros2012emotional}
Gros C.
\newblock Emotional Control--Conditio Sine Qua Non for Advanced Artificial
  Intelligences?
\newblock Philosophy and Theory of Artificial Intelligence. 2012;5:187.

\bibitem{brito2016nonlinear}
Brito CS, Gerstner W.
\newblock Nonlinear Hebbian learning as a unifying principle in receptive field
  formation.
\newblock PLoS computational biology. 2016;12(9):e1005070.

\bibitem{echeveste2015fisher}
Echeveste R, Eckmann S, Gros C.
\newblock The fisher information as a neural guiding principle for independent
  component analysis.
\newblock Entropy. 2015;17(6):3838--3856.

\bibitem{echeveste2014generating}
Echeveste R, Gros C.
\newblock Generating functionals for computational intelligence: the Fisher
  information as an objective function for self-limiting Hebbian learning
  rules.
\newblock Frontiers in Robotics and AI. 2014;1:1.

\bibitem{gros2007neural}
Gros C.
\newblock Neural networks with transient state dynamics.
\newblock New Journal of Physics. 2007;9(4):109.

\bibitem{gros2009cognitive}
Gros C.
\newblock Cognitive computation with autonomously active neural networks: an
  emerging field.
\newblock Cognitive Computation. 2009;1(1):77--90.

\bibitem{kelso2012multistability}
Kelso JS.
\newblock Multistability and metastability: understanding dynamic coordination
  in the brain.
\newblock Phil Trans R Soc B. 2012;367(1591):906--918.

\bibitem{aguilera2016extended}
Aguilera M, Bedia MG, Barandiaran XE.
\newblock Extended Neural Metastability in an Embodied Model of Sensorimotor
  Coupling.
\newblock Frontiers in systems neuroscience. 2016;10:76.

\bibitem{bekey2005autonomous}
Bekey GA.
\newblock {Autonomous robots : from biological inspiration to implementation
  and control}.
\newblock MIT Press; 2005.

\end{thebibliography}

\section*{Supporting information}


\paragraph*{S1 Video.}
\label{S1_Video}
{\bf Video for ED dissipation dynamics.} Illustrating video for
Fig.~\ref{fig_distance_probabilities}. For $n=3$ moving objects, 
a maximal object velocity of 0.1 and the ED dissipation dynamics,
as defined by (\ref{f_U_ED}).

\paragraph*{S2 Video.}
\label{S2_Video}
{\bf Video for TP dissipation dynamics.} Illustrating video for
Fig.~\ref{fig_distance_probabilities}. For $n=3$ moving objects, 
a maximal object velocity of 0.1 and the TP dissipation dynamics,
as defined by (\ref{f_U_TP}).

\paragraph*{S3 Video.}
\label{S3_Video}
{\bf Video for AT dissipation dynamics.} Illustrating video for
Fig.~\ref{fig_distance_probabilities}. For $n=3$ moving objects, 
a maximal object velocity of 0.1 and the AT dissipation dynamics,
as defined by (\ref{f_U_AT}).

\paragraph*{S4 Video.}
\label{S4_Video}
{\bf Video for ED dissipation dynamics.} Illustrating video for
Fig.~\ref{fig_timeline_ED_special}. For $n=3$ moving objects, 
a maximal object velocity of 0.5 and the ED dissipation dynamics,
as defined by (\ref{f_U_ED}).

\paragraph*{S5 Video.}
\label{S5_Video}
{\bf Video for ED dissipation dynamics.} Illustrating video for
Fig.~\ref{fig_timeline_ED_special}. For $n=8$ moving objects, 
a maximal object velocity of 0.1 and the ED dissipation dynamics,
as defined by (\ref{f_U_ED}).

\paragraph*{S6 Video.}
\label{S6_Video}
{\bf Video for ED dissipation dynamics.} Illustrating video for
Fig.~\ref{fig_timeline_ED_special}. For $n=8$ moving objects, 
a maximal object velocity of 0.5 and the ED dissipation dynamics,
as defined by (\ref{f_U_ED}).

\end{document}